\documentclass[10pt,twocolumn,preprint]{article}
\usepackage[margin=0.9in,columnse p=0.5cm]{geometry}

\usepackage[utf8]{inputenc}
\usepackage{pst-node,graphicx}

\usepackage[export]{adjustbox}
\usepackage{float}
\usepackage{amsmath}
\usepackage{authblk}
\usepackage{lineno}
\usepackage{multirow}
\usepackage{array}
\usepackage{makecell}
\usepackage{caption}
\usepackage{amsfonts}
\usepackage{enumerate}
\renewcommand\fbox{\fcolorbox{black}{blue!15!white}}
\usepackage{newfloat}
\DeclareFloatingEnvironment[fileext=boxes,name=Box]{boxes}

\title{Noise-induced Effects in Collective Dynamics and Inferring Local Interactions from Data}

\author[*]{Jitesh Jhawar}
\author[*]{Vishwesha Guttal}
\affil[*]{Centre for Ecological Sciences, Indian Institute of Science, Bengaluru, 560012}
\begin{document}
%\linenumbers
\twocolumn[{
    \begin{@twocolumnfalse}

     \maketitle

     \begin{abstract} 
       In animal groups, individual decisions are best characterised by probabilistic rules. Furthermore, animals of many species live in small groups. Probabilistic interactions among small numbers of individuals lead to a so called {\it intrinsic noise} at the group level. Theory predicts that the strength of intrinsic noise is not a constant but often depends on the collective state of the group; hence, it is also called a {\it state-dependent noise} or a {\it multiplicative noise}. Surprisingly, such noise may produce collective order. However, only a few empirical studies on collective behaviour have paid attention to such effects due to the lack of methods that enable us to connect data with theory. Here, we demonstrate a method to characterise the role of stochasticity directly from high-resolution time-series data of collective dynamics. We do this by employing two well-studied individual-based toy models of collective behaviour. We argue that the group-level noise may encode important information about the underlying processes at the individual scale. In summary, we describe a method that enables us to establish connections between empirical data of animal (or cellular) collectives with the phenomenon of noise-induced states, a field that is otherwise largely limited to the theoretical literature. %This work would be of interest to a wide range of researchers working both in theoretical and empirical analyses of collective dynamics at both cellular and organismal levels.
       \vspace{2mm} 
       \vspace{3mm}
     \end{abstract}

  \end{@twocolumnfalse}
  }]

Collective behaviour is an emergent property arising from repeated local interactions among organisms~\cite{ballerini2008interaction,lukeman2010inferring,gautrais2012deciphering,herbert2016JEB}. A number of empirical studies over the last decade have offered us novel insights on the challenging problems of characterising collective motion and on inferring underlying local interactions~\cite{katz2011inferring,herbert2011inferring,hinz2017PNAS}. Much of this success has been possible due to the availability of high-resolution spatiotemporal data of animal groups in motion, and thus in being able to reconstruct fine-scale movement of organisms. However, many of the studies only consider the {\it average or mean} properties of the group, for example average group polarisation or average degree of consensus among group members. Consequently, these studies inadvertently ignore variability of group properties, or more broadly the role of stochasticity. The conventional wisdom dictates that stochasticity often destroys order. However, this is not always the case; stochasticity may sometimes create counter-intuitive phenomena in complex systems~\cite{horsthemke1984book,ridolfi2011noise_book,boettiger2018noise} and thus deserves careful attention both in theoretical and empirical studies. %which is often thought to destroy order. 

Stochasticity in collective behaviour arises from a number of factors. Here, focusing only on factors internal to the system, we note that organisms' decisions are likely to be inherently probabilistic, either when acting on their own or when interacting with other organisms. Additionally, animal groups are finite in size and in many taxa, groups are often relatively small. In such systems, the resulting group-level stochasticity, also called the {\it intrinsic noise}, can produce nontrivial collective dynamics~\cite{horsthemke1984book, jhawar2019deriving,biancalani2014PRL,kolpas2008thesis}. 

We illustrate this concept with a simple example. Consider a colony of ants choosing between two equally good nests~\cite{biancalani2014PRL}. Assume a simple scenario in which each ant may either pick one of the two nests randomly or copy the nest choice of a randomly chosen ant. Clearly, there is no preference for ants to pick one nest over the other. We may therefore expect that ant colony members will be divided equally between the two choices and hence fail to arrive at a consensus. However, such an expectation is true only when the colony size is very large, formally called the {\it deterministic limit}. Theory predicts that if we account for stochasticity in the system, the colony does reach a consensus but only when the colony size is smaller than a threshold value~\cite{biancalani2014PRL}. This consensus is possible, intriguingly, because smaller groups exhibit more fluctuations. Therefore, in the physics literature, the collective order or consensus in this simple system is also called {\it (intrinsic-) noise-induced order}~\cite{biancalani2014PRL,jhawar2019deriving}. 

The literature on noise-induced collective behaviour is relatively small and remains largely theoretical. Apart from a recent work which demonstrates that schooling in fish is a noise-induced state~\cite{jhawar2019schooling_arxiv}, empirical work analysing stochasticity and its role in shaping collective behaviour remains at the margins of collective behaviour research~\cite{yates2009pnas,romanczuk2012mean,bertin2013njp,dyson2015,bruckner2019NatPhys}. Given that many animals live in small groups and that behavioural interactions are  inherently stochastic, we assert there is a vast scope for applying these intriguing theoretical ideas to empirical research on collective behaviour.  

In this article, we describe a method to characterise intrinsic noise in collective dynamics of animal groups~\cite{yates2009pnas,dyson2015,bruckner2019NatPhys}. We argue how such an analysis may also help us reveal local interactions that underlie the emergent patterns of collectives. The method can be applied to a highly resolved time-series of the collective state of interest; for example, the collective state could be group polarisation (or group consensus) which quantifies the degree of directional alignment (or agreement among many choices) among group members. The method we describe can be traced to van Kampen~\cite{van1981,gardiner2009} in the general context of stochastic processes in physics and chemistry but was later developed further~\cite{gradivsek2000analysis,kolpas2008thesis} and applied even in some biological studies~\cite{ghasemi2006statistical,kolpas2007pnas,yates2009pnas,boedeker2010quantitative}. However, many important issues about the method -- especially the appropriate time scale needed to characterise such dynamics -- although crucial, remain unresolved. Here, we not only aim to address such methodological issues, but also open up the potential role of stochasticity in collective dynamics of biological systems. 
\section{Noise-induced collective behaviour - a brief introduction} \label{sec:noise_ind_intro} 
In the field of collective behaviour, we are interested in how individual-level interactions (which are often stochastic) scale to emergent collective properties. To understand the role of noise in collectives, we employ the so-called {\it mesoscopic} models; this refers to a description of collective dynamics at an intermediate scale whilst explicitly accounting for the finite size ($N$) of the groups. At this (group-level) scale, probabilistic interactions produce a mean effect on the dynamics of a collective state. In addition, due to finite size of groups, there could be substantial variations (or `noise') around the mean effect. Typically, noise is expected to merely create fluctuations around the mean (e.g. a Gaussian distribution). However, when such {\it group-level noise creates new states} in the system, they are called noise-induced states~\cite{horsthemke1984book}.

In formal terms, mesoscopic dynamics of a collective state of the group, denoted by $m$, may be written in terms of a stochastic differential equation (SDE)~\cite{biancalani2014PRL,dyson2015,jhawar2019deriving}. 
\begin{equation}\label{eq:sde}
    \dot{m} = f(m) + g(m) \eta(t)
\end{equation}
(understood according to the It\^{o} convention~\cite{gardiner2009}) where $\eta(t)$ represents the noise term. Some key assumptions of this framework are:    
\begin{enumerate}[(i)]
    \item The underlying process is Markovian, i.e. the current state depends only on the previous state but not how the previous state was reached. %
    \item The noise $\eta(t)$ is uncorrelated and follows a Gaussian distribution with mean zero and unit variance ($\langle \eta(t) \rangle = 0; \langle \eta(t) \eta(t') \rangle = \delta(t-t') $). 
    \item The mesoscopic state of the collective can be described via a single coarse-grained dynamical variable ($m$).
    \item The spatial extent of the group is sufficiently small or the system is fully connected.  
\end{enumerate}

In the Discussion section, we will revisit the above assumptions of our framework in the context of applications to real systems. 

In equation~\eqref{eq:sde}, the first term $f(m)$, or more generally the {\it deterministic term}, arises from the mean effect of individual-level probabilistic interactions among group members. On the other hand, the {\it stochastic term} $g(m)$ is a consequence of variations, typically due to finite size of the system, around this mean. 

In very large groups ($N \to \infty$), the stochastic term can be ignored and only the deterministic term $f(m)$ drives the collective dynamics. In this limit, also called the thermodynamic limit in the physics literature, collective states are given by the stable fixed points of the ordinary differential equation $\dot{m} = f(m)$. 

For finite-sized groups, however, the stochastic term $g(m)$ is proportional to $1/\sqrt{N}$; thus, the strength of stochastic term is not negligible for smaller groups. We say that a system exhibits a {\it noise-induced state} when the dynamics of the finite-sized collective is qualitatively different from its deterministic limit (box 1). 

We emphasize that the noise in the SDE (equation~\eqref{eq:sde}) is at the mesoscopic or group-level. Therefore, a noise-induced state refers to a nontrivial state arising from group-level noise. Furthermore, the noise-induced state is not merely a spread/fluctuations observed around the deterministic stable state but is a new state that is absent in the deterministic limit (see box 1 for an example).  Further, a mere presence of noise at the level of individuals need not create a noise-induced state. It is often an interplay of deterministic and stochastic terms at the group-level that creates a noise-induced state. In the context of collective behaviour, if a group-level noise ($g(m)$) creates order (e.g. collective motion or consensus) that was absent in the deterministic limit, we say the system exhibits {\it noise-induced order}. 

We demonstrate these principles using two simple individual-based non-spatial stochastic models of collective behaviour from the literature~\cite{biancalani2014PRL,dyson2015}. Here, individual rules are described via stochastic interaction rates (or probabilities). The models we have chosen have contrasting collective properties, with the collective order being driven {\it stochastically} (or i.e. `noise-induced order') in one model whereas it is being driven {\it deterministically} in the other model. 

\subsection{Individual-based models of binary choice and their mesoscopic descriptions} \label{sec:ind-models}
We consider a simple scenario of decision making in a binary choice setup. Binary choices, for example, can be used to represent the nest (or food) choice of ants, as described in the Introduction section. We emphasise that we have deliberately chosen a simple framework  -- a nonspatial system with only two states -- for our study since our intention is to highlight the key principles of noise-induced states and to demonstrate a method on how to infer noise-induced states from data. Despite the simplicity, the model can be applied to contexts of decision making and even collective motion - for example, a group moving in an annulus. Indeed, this model and its extensions have been applied in a wide range of contexts such as marching locusts~\cite{dyson2012macroscopic}, fish schooling~\cite{jhawar2019schooling_arxiv}, decision making in animals~\cite{seeley2012science,marshall2019plos} recruitment of cell signaling molecules~\cite{altschuler2008spontaneous} and even financial markets~\cite{kirman1993quartecon,alfarano2008econdyn}.

Here, each individual of a group of finite size $N$ can be in one of the two states $X_1$ or $X_2$, representing their choice of the nest 1 or 2, respectively.  We denote the proportion of group members choosing nest $i$ ($i=1,2$) as $x_i=N_i/N$, where $N_i$ is simply the number of individuals choosing the nest $i$. 

The collective state of interest (also termed the {\it order parameter}) is the degree of consensus among group members defined as $m=x_1 - x_2$. A high degree of consensus (collective order) corresponds to either $m=\pm 1$. The disordered state, in which group members are split between two nests and hence do not arrive at a consensus, corresponds to $m=0$. 

We now define two models in which group members attempt to arrive at the consensus via different sets of microscopic rules. 

{\it Pairwise copying model}: In this model, individuals update their states via two mechanisms. First is a {\it spontaneous switching} where individuals change their state randomly, i.e. with no interactions with other group members, at a rate $r_1$. Using the notation of chemical reactions, this may be written as
\begin{subequations}
    \begin{equation}
        X_1 \xrightarrow {r_1} X_2,
        \label{eq:spontaneousA}
    \end{equation}
    \begin{equation}
        X_2 \xrightarrow {r_1} X_1,
        \label{eq:spontaneousB}
    \end{equation}
    \label{eq:spontaneous}
\end{subequations}
\noindent showing that spontaneous switching is unbiased. Second is the {\it pairwise copying interaction}, where a focal individual, at a rate $r_2$, copies the state of a randomly chosen individual from the rest of the group. In terms of chemical reactions, this may be written as
\begin{subequations}
	\begin{equation}
        X_1 + X_2 \xrightarrow{r_2} 2X_1,
        \label{eq:copyingA}
    \end{equation}
    \begin{equation}
        X_2 + X_1 \xrightarrow{r_2} 2X_2,
        \label{eq:copyingB}
    \end{equation}
    \label{eq:copying}
\end{subequations}
which appears to create consensus among individuals~\cite{cox1986diffusive,cox1989coalescing,granovsky1995noisy,dall2007effective,liggett2013stochastic}, but nevertheless remains unbiased between two choices. 

With these individual-level probabilistic rules, we now turn our attention to the dynamics of the collective, which is the degree of consensus ($m$) among individuals for this model. One approach to investigating collective behaviour in these models is via computer simulations of the above probabilistic rules. However, as discussed earlier, we need the analytical framework of SDEs to decipher the role of stochasticity (box 1). Recall that this in turn requires a mesoscopic description of collective dynamics, which accounts for both probabilistic interactions and finite group sizes, via stochastic differential equations. We refer the mathematically inclined readers to~\cite{romanczuk2012mean,bertin2013njp,biancalani2014PRL,dyson2015} (also see \cite{jhawar2019deriving} for a pedagogical review) for further details on deriving mesoscopic models of collective behaviour.
%
%BOX 1 
\begin{boxes*}[!]
\setlength{\fboxsep}{11pt}%%%
\fbox{\begin{minipage}{\dimexpr \textwidth\fboxsep-10\fboxrule}
\abovecaptionskip=0pt%
\caption{{\bf Noise-induced Order}}
\sbox0{\includegraphics[scale=0.4,right]{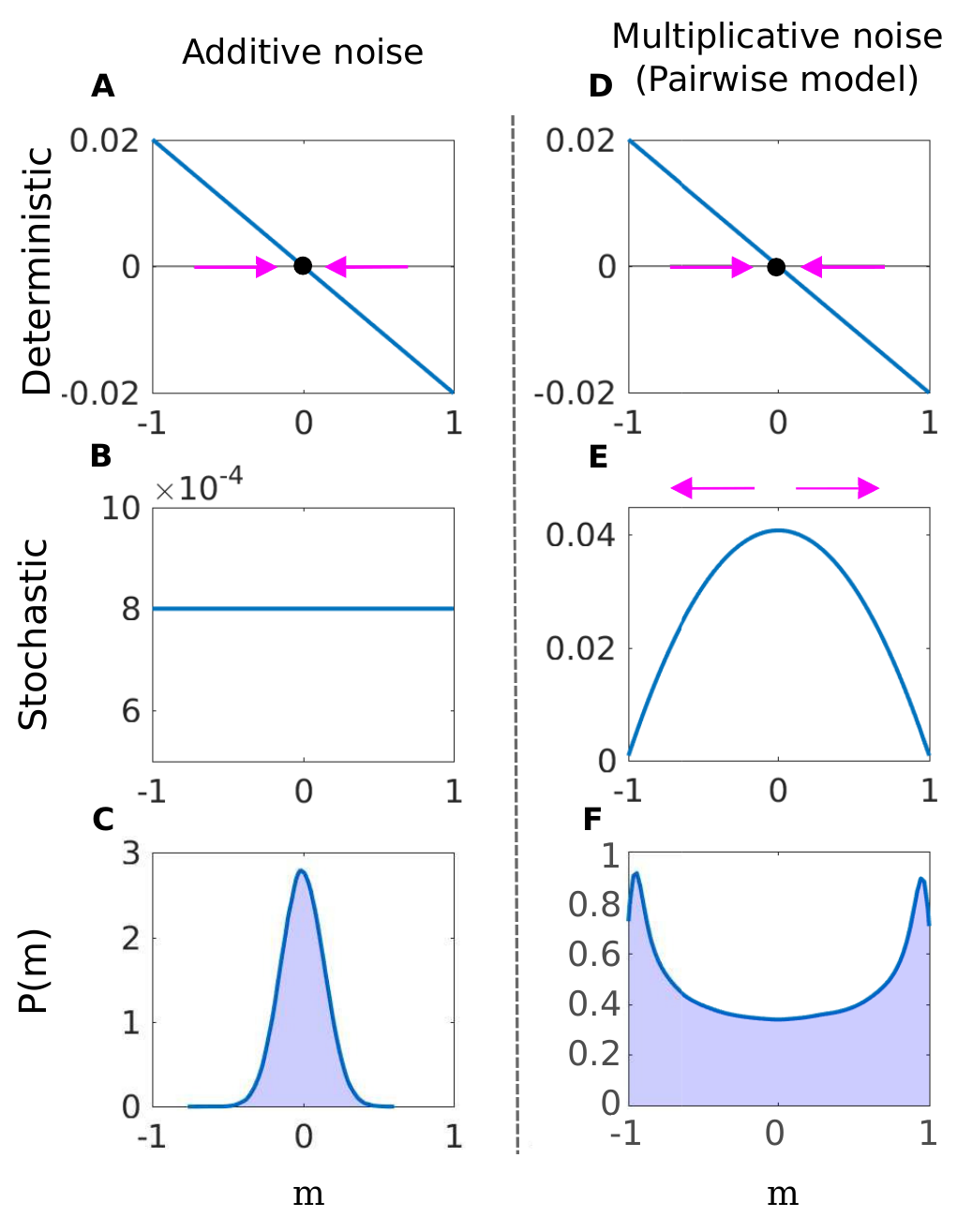}\hspace{-7cm}\vspace{5cm}\parbox{0.41\textwidth}{%
\vspace{5cm}
\captionof{figure}{An example of how noise can create new ordered states. In the first column, corresponding to equation~\eqref{eq:box_additive}, stochasticity merely creates a distribution around the deterministic stable state of disorder ($m*=0$). In the right column, corresponding to equation~\eqref{eq:box_multiplicative}, stochasticity creates new states, i.e. modes or most likely states, around $m=\pm 1$. \label{fig:sde_box}}
}}% get height and width
\hangafter -25%\numexpr -\ht0/\baselineskip -1\relax% adds two blank lines at bottom (adds blank lines before indentation)
\hangindent -7.6cm%\dimexpr \wd0 + 1em\relax% adds 1em margin on right (width of right indentation has to be negative)
\vbox{\raisebox{-\height}[10pt][0pt]{\box0}}\vspace{-2ex}% overlap image
Consider $m$ to be a quantitative descriptor of collective order, such as degree of consensus or polarisation among group members, as described in section~\ref{sec:ind-models}. Consider a hypothetical dynamic of $m$ given by the stochastic differential equation
\begin{equation}
   \dot{m} = -\alpha m + \sigma\,\eta(t),
   \tag{1.1}
   \label{eq:box_additive}
\end{equation}
where $\alpha$ is a constant and $\eta(t)$ represents uncorrelated Gaussian noise with mean zero (i.e. $\langle\eta(t)\rangle = 0$ and $\langle\eta(t)\eta(t')\rangle = \delta(t-t')$ where $\delta(t)$ is a Dirac-delta function) and $\sigma$ is the strength of the noise. This equation is also known as the Langevin equation or Ornstein-Uhlenbeck process. \vspace{0.25cm} \\  In the deterministic limit, i.e. $\sigma = 0$, the only fixed point of the system is $m^*=0$ (i.e. disorder) and is stable (figure~\ref{fig:sde_box}A). When $\sigma$ is a nonzero constant, i.e. independent of $m$ (figure~\ref{fig:sde_box}B), it is referred to as {\it the additive noise}.  Perturbations to the fixed point arising from the additive noise term are damped because the deterministic term pulls the system back to the fixed point (i.e. $m^* = 0$). Therefore, for all finite $\sigma$, the steady state probability density function $\mathcal{P}(m)$ shows a mode at zero with a width proportional to $\sigma$ (figure~\ref{fig:sde_box}C). In other words, the additive noise plays the expected role of merely `adding noise' to the deterministic stable state. \vspace{0.25cm}\\ Let us now consider the dynamic of the collective state given by the stochastic differential equation~\cite{biancalani2014PRL,jhawar2019deriving,jhawar2019schooling_arxiv},
\begin{equation}
    \dot{m} = -\alpha m + \sigma\sqrt{\alpha + \beta(1-m^2)}\,\eta(t).
    \tag{1.2}
    \label{eq:box_multiplicative}
\end{equation}
where $\alpha$ and $\beta$ are constants. This equation is inspired by the mesoscopic dynamics of the pairwise copying model introduced in section~\ref{sec:ind-models}. Here, the deterministic part is identical to that of equation~\ref{eq:box_additive} (figure~\ref{fig:sde_box}D) and hence pulls the system towards disorder ($m=0$). However, the strength of the noise depends on the current value of the state ($m(t)$) and is also referred to as {\it state-dependent} or {\it multiplicative noise} (figure~\ref{fig:sde_box}E). \vspace{0.25cm} \\
Here, when the system approaches the deterministic stable state of $m=0$, the noise strength is highest and thus, pushes the system away from disorder, $m=0$. Consequently, when $\sigma$ is above a threshold value, the most likely states of the system are in the proximity of $m = \pm 1$ (figure~\ref{fig:sde_box}F). These new most likely states in the probability density function $\mathcal{P}(m)$, which were absent in the deterministic limit, are called noise-induced states. In this case where $m$ refers to collective dynamics, we refer to the most likely states ($m= \pm 1$, corresponding to a consensus or group order) as the {\it noise-induced order}. 
\end{minipage}}
\label{Box:noise_induced}
\end{boxes*}

For the pairwise copying model, the mesoscopic dynamics of $m$ follows the stochastic differential equation~\cite{biancalani2014PRL}
\begin{equation} \label{eq:mesopairwise}
    \frac{dm}{dt} = - 2 r_1 m + \frac{1}{\sqrt{N}}\sqrt{2r_1 + (1 - m^2)r_2} \,\, \eta(t),
\end{equation}
where $\eta(t)$ represents uncorrelated Gaussian noise. 

In this SDE, the first term captures how the dynamics of consensus is shaped deterministically (i.e. mean effect), in a putative $N \to \infty$ limit. The second term captures the residual stochasticity associated with behaviour of the finite group size. The above equation can be solved analytically to obtain the steady-state probability density function of $m$~\cite{biancalani2014PRL}. Here, we focus on the intuition of dynamics driven by the above two terms.

In the limit of large group sizes ($N \to \infty$) where the stochastic term becomes negligible, the dynamics of order is given by $\dot{m} = - \alpha m$. This is a simple and well known differential equation whose stable solution is $m^*=0$. Any perturbation $|m|>0$ decays exponentially to $m=0$. In other words, any degree of consensus ($|m|>0$) will quickly decay ($|m| \to 0$) and the system becomes disordered. Hence, the deterministic (or the large group size) limit of the system does not admit consensus within groups. 

By contrast, for small group sizes the magnitude of the stochastic term -- given by $\frac{1}{\sqrt{N}}\sqrt{(2r_1 + r_2(1-m^2))}$ -- is not negligible. Moreover, stochasticity is maximum when the group is disordered ($m=0$) while it is least when there is consensus ($|m|=1$). Consequently, when $N$ is sufficiently small, stochasticity pushes the system away from the disordered state at a rate that is larger than the rate of deterministic pull towards disorder. Thus, the system achieves consensus ($|m|=1$).  

In other words, in the pairwise copying model, a curious interplay of deterministic and stochastic terms maintains order or consensus in small groups. Such a group consensus or collective order, which arises from stochasticity and is away from deterministic stable state, is termed {\it noise-induced order}. 

{\it Ternary interactions model}: In this model, individuals continue to exhibit a spontaneous switching between states at a rate $r_1$ and a pairwise copying interaction at a rate $r_2$, exactly as in Eqs~\eqref{eq:spontaneous} and~\eqref{eq:copying}. In addition, individuals exhibit a ternary interaction given by the following reactions:
\begin{subequations}
	\begin{equation}
    	2X_1 + X_2 \xrightarrow{r_3} 3X_1,
        \label{eq:ternaryA}
    \end{equation}
    \begin{equation}
        X_1 + 2X_2 \xrightarrow{r_3} 3X_2.
        \label{eq:ternaryB}
    \end{equation}
	\label{eq:ternary}
\end{subequations}
Here, interactions can happen between three individuals at a time. In an interacting triad, the individual who is in a minority switches his/her state to those of majority, at a rate $r_3$~\cite{schulze2005monte}. 

The mesoscopic dynamics of $m$ for this model is given by~\cite{dyson2015,jhawar2019deriving}
\begin{eqnarray}
\label{eq:mesoternary}
    \frac{dm}{dt} & = & - 2 r_1 m + \frac{r_3}{2}m(1-m^2) + \nonumber \\ & & \frac{1}{\sqrt{N}}\sqrt{2r_1 + (r_2 + \frac{r_3}{2})(1 - m^2)} \,\, \eta(t),
\end{eqnarray}
where $\eta(t)$ again represents the uncorrelated Gaussian noise. 

The functional form of stochastic term here is similar to that of the pairwise copying model but with an additional term associated with the ternary interaction rate ($r_3$). However, in contrast to the pairwise copying model, the deterministic term here is a cubic function arising solely from the ternary interaction rate $r_3$. Focusing on the limit of large group sizes $N \to \infty$ and thus ignoring the stochastic term, the dynamics of order is determined by the cubic equation $\dot{m} = - 2 r_1 m + \frac{r_3}{2} m (1-m^2)$. Here, when $r_3 > 4 r_1$, the system has two additional fixed points at $|m^*| > 0$ and are stable. Furthermore, the $m^* = 0$ fixed point becomes unstable.  In other words, ternary interaction model exhibits order or group consensus primarily via deterministic term. Therefore the order is present even in the large group size limit ($N\to \infty$). 

{\it Summary of model results:} In mathematical terms, the consensus in the ternary model is driven by deterministic terms and hence is realised in the large group size limit. This is a significant contrast to the pairwise model which shows consensus only when $N$ is less than a threshold value and does not admit consensus in the $N\to \infty$ limit. Therefore, the mechanism causing order in ternary interaction model is fundamentally different from the pairwise copying model, thus providing a useful contrast. 

Intriguingly, the stochastic terms are of the same form in both models, yet the importance and the role of noise is different in both models. It is also worth emphasising that an additive noise of the form shown in equation1.1 of box 1 does not produce any nontrivial ordering effects. Therefore, an interplay of deterministic and stochastic terms is necessary to produce noise-induced order.  We refer the readers to Box~2 and Table~\ref{tab:modelDetails} for an intuitive discussion of how individual-level interactions scale to mesoscale dynamical terms. 

Many previous studies on collective behaviour investigate the role of fluctuations in the collectives via an additive noise to the deterministic term of the ternary interaction model~\cite{toner1995long,solon2015prl,ramaswamy2010annrev}. In~~\cite{romanczuk2012mean}, authors begin from microscopic interactions and show that the fluctuations ({\it temperature}) depend on the group polarisation ({\it velocity field}). This implies that the state dependent (multiplicative noise) can emerge even though the individual rules have no such multiplicative noise -- which is consistent with the models we have described here~\cite{jhawar2019deriving}. However, these conclusions are derived considering the infinite-size limit or that they don't account for finite system sizes explicitly. Therefore, we reemphasise here that our framework accounts explicitly for fluctuations arising from finite-system sizes.
\begin{boxes*}[!]
\setlength{\fboxsep}{12pt}%%%
\fbox{\begin{minipage}{\dimexpr \textwidth\fboxsep\fboxrule}
\abovecaptionskip=0pt%
\caption{{\bf Linking individual-level probabilistic rules to group-level dynamics}}
\vspace{2ex}
Individual animal interactions and decisions are best modelled as probabilistic. It is not always obvious how these individual-level probabilistic interactions scale to the group-level or the mesoscopic dynamics. To understand this, recall that while the deterministic term in the mesoscopic SDE is a mean-effect of interactions, the stochastic term captures the residual variations around the mean. We now discuss these in the contexts of pairwise and ternary copying \vspace{1ex} models. 

{\bf Spontaneous switching}: The spontaneous switching of states ($r_1$) are random changes in individuals' state, without interaction with any other individuals. At the group level, the mean effect of such random state-changes is to reduce the order or consensus within groups (captured by the term $-2r_1m$ in the deterministic term of equation~\eqref{eq:mesopairwise}). As expected, individual level randomness also leads to stochasticity at the group-level ($2r_1$ in the stochastic term of \vspace{1ex} equation~\eqref{eq:mesopairwise}).

{\bf Pairwise copying interactions:} The pairwise copying interaction rate ($r_2$), surprisingly, does not appear in the deterministic term of the group-level dynamics. This is because the pairwise interactions exhibit no bias in the directionality of state-change and thus, on an average, cause equal number of individuals to switch states from $1$ to $2$ and $2$ to $1$.

However, sampling errors while individuals choose copying partners can cause substantial variation around this zero mean effect. Its effect is larger for smaller groups. When the group is at the disordered state ($m=0$), the sampling error can only cause the degree of consensus to increase and hence, the strength of noise is maximum when $m=0$. On the other hand, copying (and associated sampling errors) will have least effect at/near the ordered state ($m=\pm1$) where nearly all individuals are in the same state. Therefore,  the net effect of sampling errors due to copying is captured by the state-dependent or multiplicative noise term $(1-m^2)r_2$ in the equation~\eqref{eq:mesopairwise}. This simple structure of the noise pushes the system away from $m=0$ and when the group has high order it resides there longer due to low levels of noise. Thus, the non-monotonic structure of group-level noise, driven by pairwise copying interactions, pushes the system away from disorder ($m=0$) and towards group consensus \vspace{1ex} ($m=\pm 1$).

{\bf Ternary interactions}: Moving onto ternary interactions (with rate $r_3$), we note that it causes the minority of the three interacting partners to switch it's state towards the majority. Consequently, its mean-effect creates an ordered state and hence appears in the deterministic term of equation~\eqref{eq:mesoternary}. The residual stochasticity is exactly like the pairwise interactions. When $r_3>4r_1$, the mean or the deterministic effects alone pushes the system away from disorder towards an ordered state, the role of noise is not important to the collective dynamics in this \vspace{1ex} model. However for $r_3<4r_1$, the model behaves qualitatively similar to the pairwise model with a cubic deterministic term.

In summary, all individual-level probabilistic interactions contribute to the noise at the group level. However, interactions whose mean effect is zero at the group level does not contribute to the deterministic dynamics.

\begin{table}[H]
\centering
\begin{adjustbox}{width=0.85\textwidth}
\begin{tabular}{llll}
\hline \hline

\textbf{\makecell{Model}} & \textbf{\makecell{Stochastic \\interaction rates}} &  \textbf{\makecell{Deterministic term \\ f(m)}} &  \textbf{\makecell{Stochastic term \\ g(m)}} \\ \hline \hline

\makecell{Pairwise copying \\ model}  & \makecell{$r_1$: spontaneous switching rate\\ $r_2$: pairwise copying rate}  & \makecell{$-2r_1m$ \\ Depends on  switching \\ but not on pairwise copying} & \makecell{$\sqrt[]{\frac{2}{N}}\,\sqrt[]{2r_1 + r_2(1-m^2)}$ \\ Depends on both rates} \\

\hline

\makecell{Ternary interaction \\ model} & \makecell{$r_1$ and $r_2$: same as above \\ $r_3$: ternary interaction rate}
& \makecell{$-2r_1m + \frac{1}{2} r_3 m(1-m^2)$ \\ Depends on switching \\ \& ternary but not on pairwise copying} 
& \makecell{$\sqrt[]{\frac{2}{N}}\,\sqrt[]{2r_1 + (r_2+\frac{r_3}{2})(1-m^2)}$ \\ Depends on all three rates} \\
%& ${\sqrt{\frac{2}{N}}\sqrt[]{2r_1+(r_2+\frac{r_3}{2})(1-m^2)}$ \\
\hline \hline
\end{tabular}
\end{adjustbox}
\caption{Scaling from individual stochastic interaction rates to group-level dynamics (deterministic and stochastic terms).}
\label{tab:modelDetails}
\end{table}
\end{minipage}}
\label{Box:scaling}
\end{boxes*}

\section{Characterising noise-induced states from data}
\label{sec:char_noise}

With this background, we now turn our attention to the inverse question - which is also the main goal of this manuscript: Given a time-series data of a collective state (or order parameter), we ask is it possible to infer if the order was noise-induced? 

\begin{figure*}[!]
    \centering
    \includegraphics[width=0.75\textwidth]{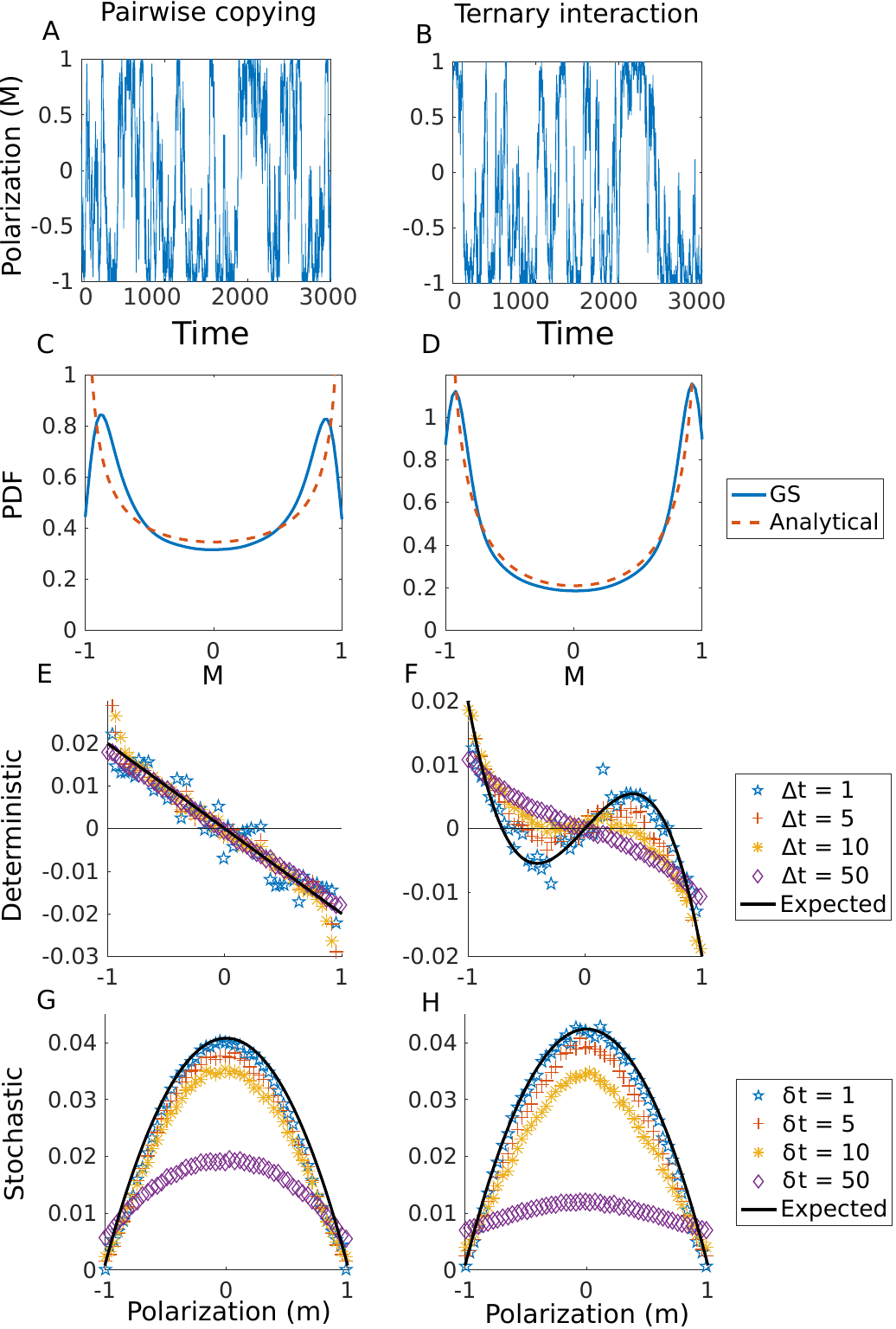}
    \caption{\textbf{Characterizing mesoscopic dynamics of  collective behaviour from data.} \textbf{[A, B]} Representative time-series of $M(t)$ for the pairwise and the ternary interaction models show that system does not reach an equilibrium value. \textbf{[C, D]} Probability density function of the data show two modes corresponding to ordered states for both the pairwise and the ternary interaction model. The red dashed line represents the PDF derived using the analytical expression (see~\cite{jhawar2019deriving}). \textbf{[E, F]} Data-derived deterministic terms match the expected functional forms when $\Delta t = 50$ for the pairwise and $\Delta t \approx 5$ for the ternary model. \textbf{[G, H]} Data-derived stochastic terms for both the models are similar and match the expected function when $\delta t = 1$. The black solid line in panels \textbf{[E - H]} represents the analytical forms. Parameters: (Pairwise model) $r_1 = 0.01$, $r_2 = 1$, $N = 50$. (Ternary model) $r_1 = 0.01$, $r_2 = 1$, $r_2 = 0.08$, $N = 50$.}
    \label{fig:data_both_models}
\end{figure*}

To address these questions, we perform stochastic simulations of both pairwise and ternary interaction models using the Gillespie algorithm~\cite{gillespie1976,gillespie1977}. In panels A and B of figure~\ref{fig:data_both_models}, we display the time-series of the degree of consensus (denoted by $M$) among 50 individuals using the pairwise and ternary copying models, respectively. We denote the order parameter obtained by simulations by the capital letter $M$. 

We observe that in both systems the degree of consensus does not reach an equilibrium value but shows dynamic patterns, sometimes reaching a consensus ($M=\pm 1$) but repeatedly switching back and forth between two consensus values (i.e., $M=1$ or -1). 

In panels C and D of figure~\ref{fig:data_both_models} we display the graphs of the probability density functions  of $M$. These show that the most likely state in both models is a high degree of consensus ($M \approx \pm 1$). 

We recall that there are fundamental differences between the nature of collective dynamics in these models; while the collective order in the pairwise copying model is driven by stochasticity (i.e., noise-induced), the order in the ternary copying system is entirely driven by the deterministic term. Yet, visual inspection reveals no qualitative features that distinguish the two model outcomes in figure~\ref{fig:data_both_models}A-D --- either in terms of dynamics or the most likely states. 

However, as shown in the previous section, the SDEs that govern the dynamics of the consensus in two models are indeed different. Therefore, if we can use the time-series data shown in figure~\ref{fig:data_both_models} panels A and B to construct SDEs of the form (see box 1)
\begin{equation}
	\dot{m} = F(m) + G(m)\eta(t),
	\label{eq:generic-sde}
\end{equation}
we may decipher the role of stochasticity in each of the datasets. Here, as before, $\eta$ is a Gaussian white noise with mean zero and unit variance, $F(m)$ represents the deterministic term (also called {\it drift} coefficient) and $G(m)$ is the stochastic term (with $G^2(m)$ called the {\it diffusion} coefficient) driving the dynamics. 

We note that we have used capital letters to denote simulated data ($M$) and the data-constructed functional forms of deterministic ($F(m)$) and stochastic terms ($G(m)$). While the simulated $M$ is necessarily discrete owing to finite number of individuals $N$ in simulations, the order parameter in SDEs is assumed/approximated as a continuous order parameter; hence in the functional form we keep the $m$ notation, resulting in a composite notation such as $F(m)$ and $G(m)$. This notation also helps to distinguish from analytically derived formal equations such as equations~\eqref{eq:mesopairwise} and \eqref{eq:mesoternary}. 
\subsection{Method for constructing SDEs from data}
Following~\cite{van1981,gradivsek2000analysis,kolpas2007pnas,yates2009pnas,friedrich2011approaching}, the deterministic component (or the drift coefficient) can be approximately obtained by the {\it first jump-moment} defined as
\begin{equation}
	F(m) = \left\langle \frac{M(t+\Delta t) - M(t)}{\Delta t}\right\rangle\Bigg\vert_{M(t) \in [m, m+\epsilon ]},
	\label{eq:deterministic}
\end{equation}
where the angular brackets denote an average over all instances in the time-series where $M(t)$ is close to a given $m$. In either real or simulated time-series, the observable will never (or rarely) be exactly equal to a given $m$; hence the average is obtained considering all $M \in [m, m +\epsilon]$, where $\epsilon$ is a small value (we choose $\epsilon=0.01$). In other words, the deterministic part $f(m)$ is the {\it average or expected change per unit time} in the observable quantity when it is at (or near) the value $m$.  

Likewise, the stochastic term (or the diffusion coefficient) can be approximately computed via the {\it second jump-moment}, defined as
\begin{subequations}
\begin{equation}
G^2(m)  = \left\langle \frac{R^2(m)}{\delta t}\right\rangle \tag{9}
\end{equation}
\begin{equation}
\begin{split}
    \textrm{where,} \,\, R(m)	& =  (M(t+\delta t) - M(t))\vert_{M(t)  \in [m, m+\epsilon ]}  \\
    & \qquad - F(m) \, \delta t \nonumber
\end{split}
\end{equation}
\label{eq:stochastic}
\end{subequations}
Here too, the averaging is done over the entire time-series, and in the vicinity of $m$ as described above. 

To obtain an intuition for this formula, we decompose the {\it residual} term $R$ into two parts: The first part is the term $M(t+\delta t) - M(t)$, representing the actual change in the observable over a time $\delta t$ from $t$. The second part is $F(m) \delta t$, which is the expected change in the observable based on the deterministic term alone. Therefore, for any given value of $m$, the term $R$ in the numerator is basically the difference between the observed change and the expected change from the deterministic term. Considering squaring of this difference and the averaging, we may readily recognise the numerator as the second moment and hence captures the stochasticity in the dynamics of the state variable $m$. 

Although this method of constructing equations from data has been used earlier~\cite{yates2009pnas,boedeker2010quantitative}, a fundamental issue of choosing the right timescales to compute the deterministic and stochastic terms is overlooked. Note that we have deliberately chosen different notations for time steps $\Delta t$ and $\delta t$ in the formulae to compute the deterministic and stochastic terms, respectively. A naive choice could be that both time steps must be equal to the smallest time step, i.e., at the finest resolution in which data is available. However, that is not the case. Here, we conjecture and later confirm via simulations that appropriate time scales for constructing the deterministic and stochastic forces are not the same. More specifically, while $\Delta t$ must be comparable to the autocorrelation time of the time-series data, $\delta t$ must be much smaller provided that the Gaussian approximation of noise in the mesoscopic SDE is still valid (see box 3). 

%MORE METHODS about optimum D/delta t figures: During revision. 
% box 3
\begin{boxes*}[!]
\setlength{\fboxsep}{9pt}%%%
\fbox{\begin{minipage}{\dimexpr \textwidth\fboxsep-10\fboxrule}
\abovecaptionskip=0pt%
\caption{{\bf Time scale to construct the mesoscopic dynamics}}
\vspace{2ex}
{\bf Time-scale to construct deterministic term: $\mathbf{\Delta t}$}: The time scale  to compute deterministic component must be comparable to (but less than) the autocorrelation time--- the time difference above which two
measurements of the observable M become essentially uncorrelated (see Appendix section~A.1
for formal definition)--- of the time-series (denoted by $\tau_c$). At very fine time scales ($\Delta t \ll \tau_c$ ), stochasticity of individual-level interactions will cause constant perturbations to the system away from deterministic stable state. The time scale over which these perturbations decay and system relaxes back to deterministic stable states is typically given by $\tau_c$. If we choose a $\Delta t \gg \tau_c$, we are likely to miss the relaxation dynamics of perturbations. Therefore, to capture the dynamics driven by the deterministic forces, we conjecture that a time scale $\Delta t$ comparable to $\tau_c$ is most appropriate. \\

{\bf Time-scale to construct stochastic term: $\mathbf{\delta t}$}: The time scale to compute stochastic component should be much smaller to capture stochastic effects provided the number of probabilistic events in that time window follow a Gaussian distribution. Equivalently, the residuals $R(m)$ for any $m$ must follow a Gaussian distribution. This expectation is based on the key assumption of the mesoscopic SDE description where the noise $\eta(t)$ is uncorrelated and follows a Gaussian distribution. For larger time window, although these assumptions could still hold true, the stochastic effects would average out. Therefore, we expect that $\delta t$ is much smaller than $\Delta t$.

\end{minipage}}
\label{Box:conjecture}
\end{boxes*}
%
%
%
%Figure 1
\subsection{Model parameters}
\label{subsec:param}
{\it Parameter values:} We apply the above described method on the data generated from simulation of the models. For generating this data we vary different parameters ($N,\,r_1, \,r_2,\,r_3$) in the models. 

The pairwise model is a special case of the ternary interaction model with  $r_3 = 0$. Without loss of generality, we set $r_2=1$. We then choose $r_1=0.01$, which as per the analytical results of~\cite{biancalani2014PRL} sets the critical system size to $N_c = 100$; that is, if $N<N_c$, the group exhibits noise-induced order ($m>0$), whereas the group exhibits disorder ($m \approx 0$) if $N>N_c$.  Therefore, for demonstration of the method, we choose three different values of $N = 50,\,100,$ and $200$ corresponding to the so-called sub-critical, critical and super-critical regimes, respectively. 

For the ternary model, when the ternary copying rate is sufficiently high (specifically, $r_3>4r_1$), the system exhibits deterministically driven ordered dynamics ($m>0$). Keeping the rest of the parameters same as the pairwise model, we arbitrarily set $r_3 = 0.08$ for our analyses. We note that the dynamics of this model is qualitatively similar to the pairwise model in the sub-critical regime of this model ($r_3 < 4 r_1$) and hence we do not consider those values for our demonstration. 

{\it Correlation time:} The correlation time ($\tau_c$) of a time series is the time difference above which two measurements of the observable become essentially uncorrelated (see Appendix section~A.1 for a formal definition). We find that $\tau_c$ does not change with $N$ for the pairwise model, but it increases with $N$ for the ternary model. For both the models, $\tau_c$ decreases with the spontaneous rate $r_1$ (electronic supplementary material figure~S2). Hence, to understand the role of correlation time in the reconstruction of the SDE, we vary $r_1$ for the pairwise model and $N$ for the ternary model.

{\it Distance between the expected and derived function:} We measure the distance between the derived and the expected functions by calculating the normalised root mean square distance metric ($D$), defined as 
\begin{equation}
D(F,f) = \frac{\| A - a \|_2}{\| a \|_2},
\label{eq:distance}
\end{equation}
where  $\displaystyle \| f \|_2 = \sqrt{\sum_m f^2(m)}$ defines the  $\ell^2$-norm for a function $f$. $A$ represents derived function and $a$ represents the expected functional form from the analytic derivations of the mesoscopic SDE. 
\subsection{SDEs constructed from data reveal the role of stochasticity in collective dynamics}
\label{sec:sde_recon}
We now demonstrate the method of SDE construction by using the data generated by individual-based collective behaviour models described in section~\ref{sec:ind-models} (see figure~\ref{fig:data_both_models}A \& B for representative graphs of time-series). 

To construct the deterministic term, we apply equation~\eqref{eq:deterministic} to time-series of $M$ for both models. For the pairwise copying model, we find that the deterministic term is a linear function of $m$ (figure~\ref{fig:data_both_models}E). Analysis of the time-series of the ternary interaction model reveals a deterministic term which is a cubic function of $m$~(figure~\ref{fig:data_both_models}E). Reassuringly, the functions thus constructed for both models match remarkably well with the analytically expected deterministic terms of equations~\eqref{eq:mesopairwise} and~\eqref{eq:mesoternary}.
%
%Figure drift distance
\begin{figure*}[t!]
\centering
\includegraphics[width=0.75\textwidth]{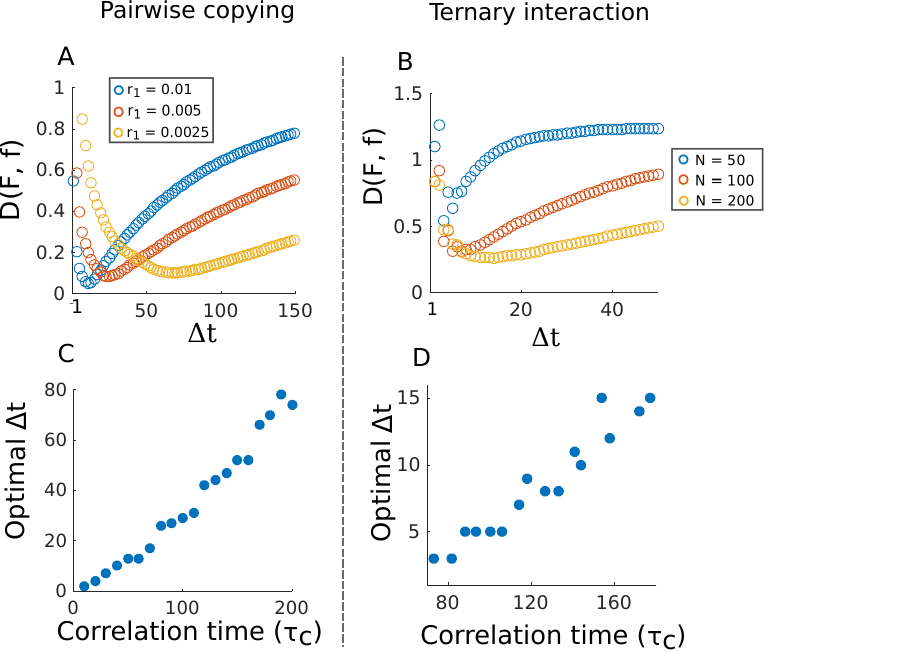}
\caption{\textbf{Optimum time scale to extract the deterministic term.} Distance between the data-derived and the expected form for the deterministic term, $D(F,f) = \frac{\| F - f \|_2}{\| f \|_2}$, is used to find the optimum time scale ($\Delta t_{opt}$) for extraction (see equation~\eqref{eq:distance}]). {\bf[A]} shows this distance as a function of $\Delta t$ for the pairwise model for three values of spontaneous reaction rate ($r_1$) and {\bf [B]} three values of system size ($N$) for the ternary model; {\bf note  that parameters of the model that influence correlation time in that model are chosen for this analyses (see section~\ref{subsec:param})}. The $\Delta t_{opt}$ corresponding to the minima of above plots is the optimum time scale to derive the deterministic component. {\bf [C} and {\bf D]} show that $\Delta t_{opt}$ as a function of correlation time ($\tau_c$) follow the same pattern for both models, suggesting a possible universal rule that  $\Delta t_{opt}$ is roughly an order of magnitude less than $\tau_c$.}
\label{fig:optimum-Dt}
\end{figure*}

In the above data-driven construction of deterministic dynamics, we considered a range of values of $\Delta t$. The results for some $\Delta t$ are shown in panels figure~\ref{fig:data_both_models}E-F. The smallest time step $(\Delta t = 1)$ yields a noisy pattern around the analytically expected functions for both the models. However, the constructed functions become closer to the analytical expectations (Table~\ref{tab:modelDetails}) for larger values of $\Delta t$. When $\Delta t$ is around an order of magnitude less than the autocorrelation time of the time-series, we find that the fit is most accurate, i.e. the distance between the analytically expected and the data constructed functions reaches a minimum value (figure~\ref{fig:optimum-Dt}A-B). Furthermore, we find a strong positive relationship between optimum value of $\Delta t$ and the autocorrelation time ($\tau_c$) (figure~\ref{fig:optimum-Dt}C-D); $\tau_c$ changes as model parameters vary (see section~\ref{subsec:param} and Appendix section~A.2)). %of the given time-series is true for a wide range of parameter values of both the pairwise and ternary interaction models (figure~\ref{fig:optimum-Dt}C-D and electronic supplementary material section~S1.2).
\begin{figure*}[t!]
\centering
\includegraphics[width=1.0\textwidth]{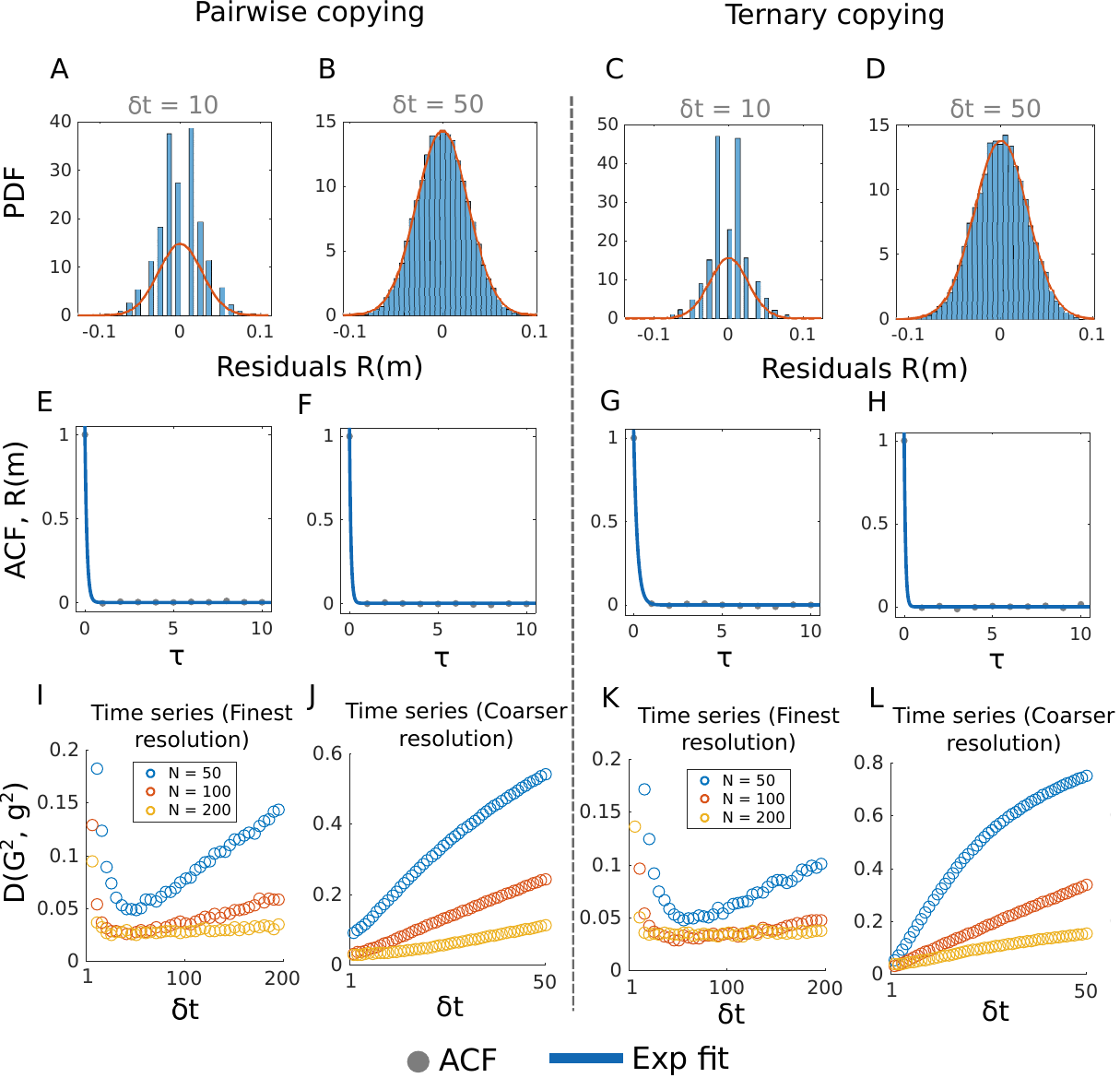}
\caption{\textbf{Optimum time scale to extract the stochastic term} {\bf (Top row)} Distribution of residuals $R(m)$ for $M=0$ for different $\delta t$ using the fine time-series for both models. For both the models, the distribution is not perfectly Gaussian for smaller $\delta t$ {\bf [A, C]} but increasing $\delta t$ improves the Gaussian approximation {\bf [B, D]}. {\bf (Middle row)} {\bf [E-H]} The autocorrelation function of the residuals ($R(m)$) for $M=0$ shows that noise is uncorrelated in our simulated datasets. {\bf (Bottom row)} Distance, $D(G,g) = \frac{\| G^2 - g^2)\|_2}{\| g^2 \|_2}$ (see equation~\eqref{eq:distance}]), plotted as a function of $\delta t$ for the pairwise copying model and the ternary interaction model for three values of system sizes ($N$). In {\bf [I, K]}, the time-series is much finer and $\delta t_{opt}>1$. In  {\bf [J]} and  {\bf [L]}, the time-series is coarse and hence we find $\delta t_{opt} = 1$.    Therefore the optimum ($\delta t_{opt}$) corresponds to the smallest $\delta t$ at which the Gaussian structure of noise is preserved.}
\label{fig:optimum-dt}
\end{figure*}

We now turn our attention to constructing the stochastic term, by applying equation~\eqref{eq:stochastic} to time-series data from both models, for a range of values of $\delta t$. Here too, for both models, we are able to obtain the analytically expected functional form of an inverted parabolic function to a remarkable accuracy. Interestingly, the smallest $\delta t$ yields the most accurate stochastic force function. This match becomes rapidly worse with increasing the time step ($\delta t$), a pattern exactly opposite to that of constructing the deterministic term (figure~\ref{fig:data_both_models} G-H). 

\begin{figure*}[t!]
\centering
\includegraphics[width=1.0\textwidth]{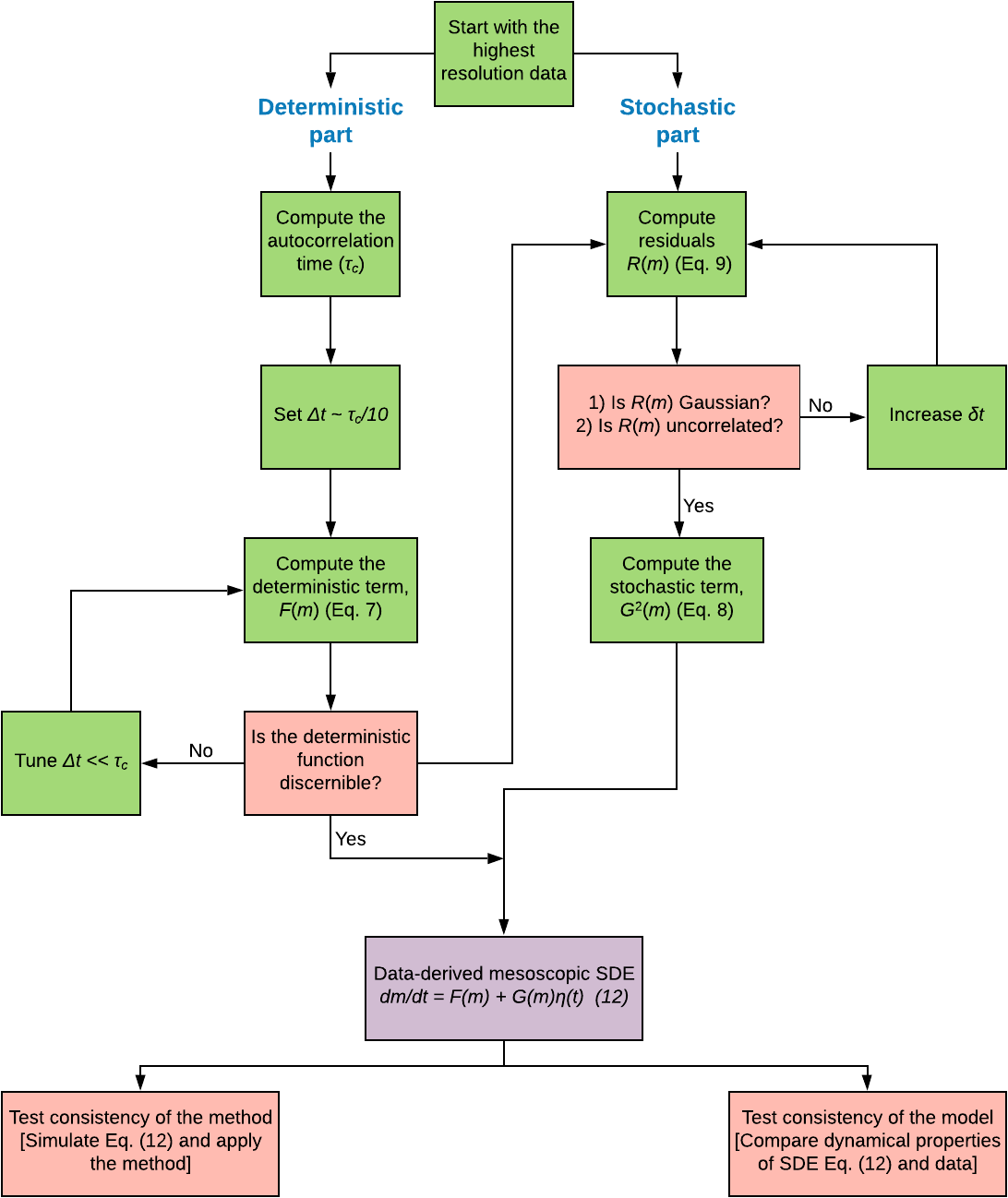}
\caption{\textbf{A flowchart summarising the method to derive the stochastic differential equation~\eqref{eq:generic-sde} from time-series data.}}
\label{fig:flowchart}
\end{figure*}

We explore the role of $\delta t$ further, by first generating very high-resolution data of both individual-based models and then analyzing the noise ($\eta (t)$) structure. Interestingly, we find that for very high resolution data the noise distribution is not perfectly Gaussian (Figure~\ref{fig:optimum-dt}A \& C), thus violating the assumption of SDE~\eqref{eq:sde}. We find the distribution tends to Gaussian for coarser time scales ( Figure~\ref{fig:optimum-dt}B \& D, also see Appendix section~D for $\delta t = 100$). We also analyze the correlations in noise by plotting its autocorrelation function. For all time scales in our analysis we find the noise in the data to be uncorrelated (figure~\ref{fig:optimum-dt}I-L), satisfying a key assumption of our framework (section~\ref{sec:noise_ind_intro}). Both these checks, i.e. whether $\eta(t)$ is Gaussian as well as uncorrelated in data are important to validate key underlying assumptions (figure~\ref{fig:flowchart}). Furthermore, though the distribution of the residuals/noise ($R(m)$) is a Gaussian above a certain time scale in the data, we conjecture and confirm that there is an optimum of $\delta t$ beyond which the extraction of stochastic term becomes inaccurate (Figure~\ref{fig:optimum-dt}I \& K). 

However, if the data was available at a coarser resolution, we may not find the optimum $\delta t$ as depicted by Figure~\ref{fig:optimum-dt}I \& K. For example, if the time interval between consecutive data points is more than the optimum $\delta t$, we find that the smallest time step of such coarser time series yields minimum error in the reconstruction (Figure~\ref{fig:optimum-dt}J \& L). Indeed, this is the reason why we find that the best reconstruction of the stochastic term in figure~\ref{fig:data_both_models} corresponds to $\delta t=1$. 

Finally, we confirm that the method of construction of SDE from simulated data of the individual-based models is consistent for different parameter values in models (Appendix section~B). Reassuringly, in all the cases, the data-derived deterministic and stochastic functions match not only the qualitative features of the analytically expected functions but also quantitatively. 

\subsection{Consistency of the method and model}
We ask whether our proposed method is {\it self-consistent}? To do this, we simulate the data-derived SDE to generate high-resolution time series of the collective state variable $m$. We then apply the same set of steps to reconstruct the SDE from this time series. We find that such a procedure yields comparable values of time scales ($\Delta t$ and $\delta t$) and qualitatively similar forms of the deterministic and stochastic terms of the SDE (Appendix section~C), thus demonstrating the consistency of the method.  

Next, we ask whether the data-derived SDE model produces dynamical features consistent with the experimental data; the latter in our case corresponds to Gillespie simulation of the microscopic interaction rules. We recall that the construction of SDEs relied only on the first and second jump-moments, together with the autocorrelation time, of the time series of the state variable. We consider two functions -- (i) autocorrelation function of $m$ and (ii) probability density function of $m$ -- neither of which were directly used to construct the model. We find that the autocorrelation function of the timeseries of the SDE model is in excellent qualitative agreement with that of the Gillespie simulations of the microscopic model (electronic supplementary material figure~S1). Likewise, the probability density function of $m$ computed from the SDE and the Gillespie simulations of the microscopic model too show qualitative agreement across a range of parameter values for both models (figure~\ref{fig:data_both_models}C-D and electronic supplementary material figure~S3). Put together, these analyses suggest self-consistency of both the method of construction of SDE that we proposed as well as the model we have derived.  
%
%
%Figure flowchart

\section{Discussion}
We demonstrate a method to characterize the dynamics of collective behaviour that accounts for intrinsic stochastic effects in groups. Such noise arises due to small sizes of groups and probabilistic interactions among group members. Specifically, given a high-resolution time-series data of collective behaviour, we characterised the dynamics via a stochastic differential equation (SDE) which accounts for both deterministic and stochastic drivers. Our key contribution lies in finding optimum time scales over which to compute the deterministic and stochastic terms of SDE. With this characterization, we highlight the potential of intrinsic noise in producing group order even though deterministic limit does not predict order. 

\subsection{Novelty and applicability of the method}
Strikingly, this method can help us distinguish whether the observed collective order is due to deterministic or stochastic drivers. To demonstrate this, we use two well studied toy models of collective behaviour. For these models we know the exact forms of the mesoscopic scale SDEs from previous analytical studies~\cite{mckane2004,biancalani2014PRL,dyson2015}. Specifically, while the pairwise interaction model exhibits a noise-induced order, the ternary interaction dominated system exhibits order driven by deterministic terms. The qualitative features of the time-series of collective behaviour for these models are similar (figure~\ref{fig:data_both_models}A-D). From the same high-resolution temporal data of the group order, we are able to confirm that the described method faithfully characterizes the mesoscopic SDEs. This is reassuring and therefore instils confidence that we can employ the method in more complex scenarios including real data. 

Although simple and elegant, this method has rarely been used in the biology literature (but see~\cite{boedeker2010quantitative,kolpas2007pnas,yates2009pnas,jhawar2019schooling_arxiv}). One possible reason might be the lack of clarity on methods of constructing the deterministic and stochastic terms. A key finding from our study is that the construction of both the parts need to be done at different time scales in the data. To construct deterministic term, a time scale slightly smaller than the autocorrelation time of the data seems optimal. In contrast,  stochastic term needs to be constructed at much finer time scales. In figure~\ref{fig:flowchart} we provide a flow chart of the procedure. 

In the context of collective behaviour, a study on marching locusts applied the same method and found an evidence for multiplicative noise which is of the similar form as the pairwise and ternary interaction models~\cite{yates2009pnas}. However, in their system, the deterministic term was cubic, like in the ternary model. Hence, the deterministic term alone could explain the order. A recent study on fish schools ({\it Etroplus suratensis}) of small to intermediate group sizes shows that the highly aligned motion is a noise-induced effect, best explained by the simple pairwise alignment interaction model we discussed above~\cite{jhawar2019schooling_arxiv}. The method has also been applied to study single cell migrations  and find that movement in normal and cancerous cell types differs qualitatively. While cancerous cells show migration that is driven purely due to deterministic effects, normal cells are driven only by stochastic factors~\cite{bruckner2019NatPhys}. These examples show that not only does the method offer a rigorous quantitative description of collective dynamics (or even single organismal behaviour), but may also offer insights on the individual-level processes.

We have demonstrated this method for two nonspatial models of collective behaviour where each individual could be in one of the two discrete decision states. Despite the simplicity of the model framework we used, the applicability of the method to construct mesoscopic dynamics is wider. For example, as in the locust study~\cite{yates2009pnas}, two states could be interpreted as two directions of movement in an annulus, and hence the group order may correspond to the degree of alignment of collective motion. For swarming systems in a continuous two (or three) dimensional world, the number of states are infinitely large. However, we may define the group order using a vectorial representation and construct a mesoscopic SDE by a fairly trivial extension of the method proposed here~\cite{jhawar2019schooling_arxiv}. %We expect that the method to construct dynamics of collective order via SDEs to be valid in these generalisations. 
\subsection{Limitations and future directions}
The method to construct mesoscopic SDEs requires further exploration in several contexts. A key assumption of the SDE equation~\eqref{eq:sde} is that the noise ($\eta(t)$) is Gaussian and uncorrelated.  However, time series (Finest resolution), in many systems noise can be temporally correlated~\cite{hanggi1978correlation} and may exhibit large fluctuations~\cite{murakami2015inherent}. Therefore, it is important to test these assumptions while applying the method to real datasets (see figure~\ref{fig:optimum-dt}~E-H). Furthermore, in the wild, the dynamics of animal groups is likely influenced by external stimuli as individuals respond to threats, food availability, mates and so on. Consequently, the resulting group dynamics may often be non-stationary. However, our method is applicable only to stationary time series. Thus, it would be interesting to generalise the method of SDE construction in such complex scenarios. 

The dynamics of other collective state variables such as rotational and dilational order are also of interest in many biological and physical contexts. Our method can be readily applied to these order parameters as well. However, in many contexts the state of a collective may not be fully described by a single coarse-grained variable but rather requires coupled dynamical variables (e.g. density together with the order parameter). Future work may extend the method of characterisation of noise to such scenarios. 

In this study, we considered only simple non-spatial descriptions. While neglecting space might be justified for small groups, in larger groups explicitly accounting for space is crucial to analyze the role of fluctuations. Recent analytical work suggests that we can derive stochastic partial differential equations to account for finite-group dynamics~\cite{laighleis2018pre,chatterjee2019three} where group order can be described as a function of both time and space. %Therefore, it is important to explore the possible ways by which the method can be extended to derive such equations directly from data of large collectives spread over space. 
We note that promising efforts have been made in the context of Navier-Stokes equations of physical systems~\cite{rudy2017data}. These approaches may shed light on extending our approaches to develop data-driven hydrodynamic descriptions of collective motion~\cite{toner1995long,ramaswamy2010annrev,grossmann2013njp}. 

Finally, we comment on the potentially interesting biological consequences of noise-induced order. A system exhibiting (intrinisic) noise-induced states is characterized by the existence of a critical group size, at which key collective properties qualitatively change. This is particularly important for the design of experiments on collective behaviour, with the implication that conclusions based on small (or large) group sizes cannot be trivially applied to other group sizes. Another important question that merits attention is whether the noise-induced collective behaviour is adaptive. Natural selection is strongest at the level of individuals; therefore, it is likely that the observed mesoscopic noise-induced order is a simple consequence of selection for the microscopic interactions (e.g. pairwise copying versus higher-order copying); however, we may expect nontrivial feedback from the emergent properties of the group. It would be interesting to explore, both theoretically and experimentally, the adaptive significance of noise-induced collective behaviour. 
%\textcolor{red}{Moreover, related theoretical studies in the context of hydrodynamic descriptions~\cite{solon2015prl}, active nematics~\cite{bertin2013njp} as well as population models and cell polarization~\cite{mckane2014stochastic} highlighting the role of space instills this point further.}
%

\subsection{Concluding remarks}
Our study highlights a much neglected but an elegant concept of noise-induced states in empirical studies on collective behaviour. The method we described to infer the role of stochasticity can be readily applied to data on collective motion across animal species. Stochastic interactions as well as finite (and small) sizes are inescapable features of biological world. Therefore, we expect that noise-induced states are likely to occur in larger classes of biological phenomena. We hope that our study inspires further studies on the role of noise in collective motion, especially with a focus on functional aspects of collective behaviour.  
\section{Declaration}
The authors declare no conflict of interest.

\section{Codes for data analyses}
The codes used for analysis in the manuscript and to analyze real data can be downloaded from https://github.com/tee-lab/Characterizing\_noise.

\section{Correspondence}
To whom correspondence should be addressed. Email: jiteshjhawar@gmail.com

\section{Acknowledgements}
We thank Richard G Morris, M Danny Raj, Aakanksha Rathore, Vivek Jadhav and Ayan Das for discussions and critical comments on the manuscript. VG acknowledges support from DBT-IISc partnership program, SERB (DST) and infrastructure support from DST-FIST. JJ acknowledges support from CSIR-India for research scholarship.

\bibliographystyle{unsrt}

\clearpage
\appendix
\renewcommand\thefigure{\thesection.\arabic{figure}}
\setcounter{figure}{0}
%\twocolumn[{
%\begin{@twocolumnfalse}

\section{Correlation time} \label{app:corr_time}
\subsection{Determining correlation time of the time series}
\begin{figure*}[!]
\centering
\includegraphics[width=0.6\textwidth]{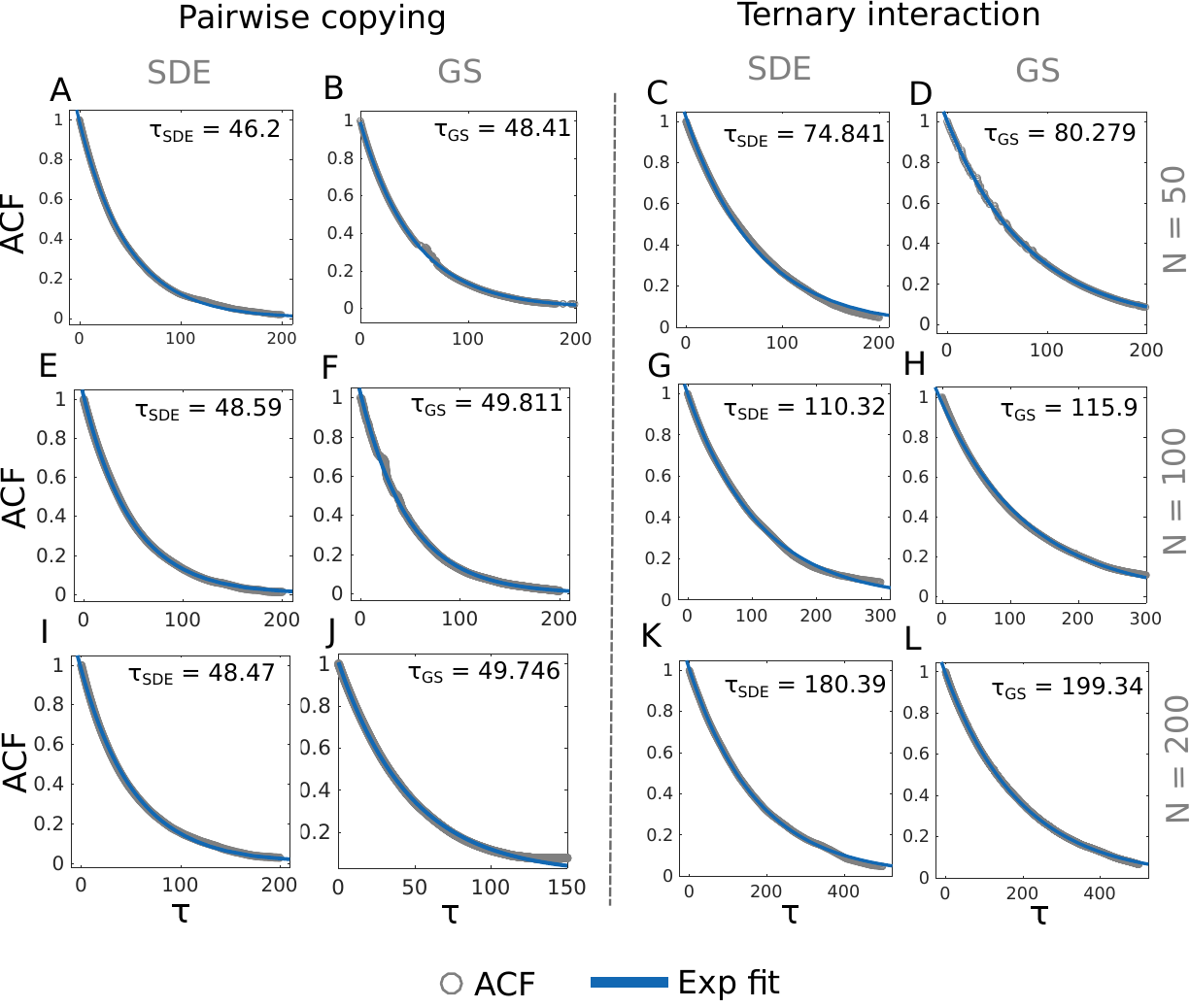}
\caption{\textbf{Autocorrelation functions (ACF)} ACF plotted as a function of $\tau$ for both the pairwise copying and the ternary interaction model from both the data from SDE as well as individual-based models. The blue line in all the plots represents the exponential fit ($ae^{-b\tau}$) to the ACF. The correlation time  is the inverse of the exponent, i.e. $\tau_c = 1/b$. The values of correlation times from both kinds of data for both models are in good agreement.}
\label{fig:corr_fit}
\end{figure*}

We used the standard formula to calculate the autocorrelation function (ACF) of the time series of the state variable $M(t)$. 
\begin{equation}
    ACF(\tau) = \frac{\langle (M(t) - \langle M(t)\rangle)(M(t+\tau) - \langle M(t)\rangle)\rangle}{\langle(M(t) - \langle M(t)\rangle)^2\rangle}.
\end{equation}
This gave us the plot of ACF as a function of time lag $\tau$ (figure~\ref{fig:corr_fit}). We then fitted an exponential function $ae^{-b\tau}$ to this plot. The correlation time ($\tau_c$) is then given by the inverse of the exponent, i.e. $1/b$. The final value of $\tau_c$, for a given parameter value that we report is an average value calculated using 100 replicates of time series data each comprising of 100,000 time steps.

% figure correlation fit plots figure 1

\subsection{Analysing correlation time for the pairwise and the ternary interaction model}
\begin{figure*}[t!]
\centering
\includegraphics[width=0.6\textwidth]{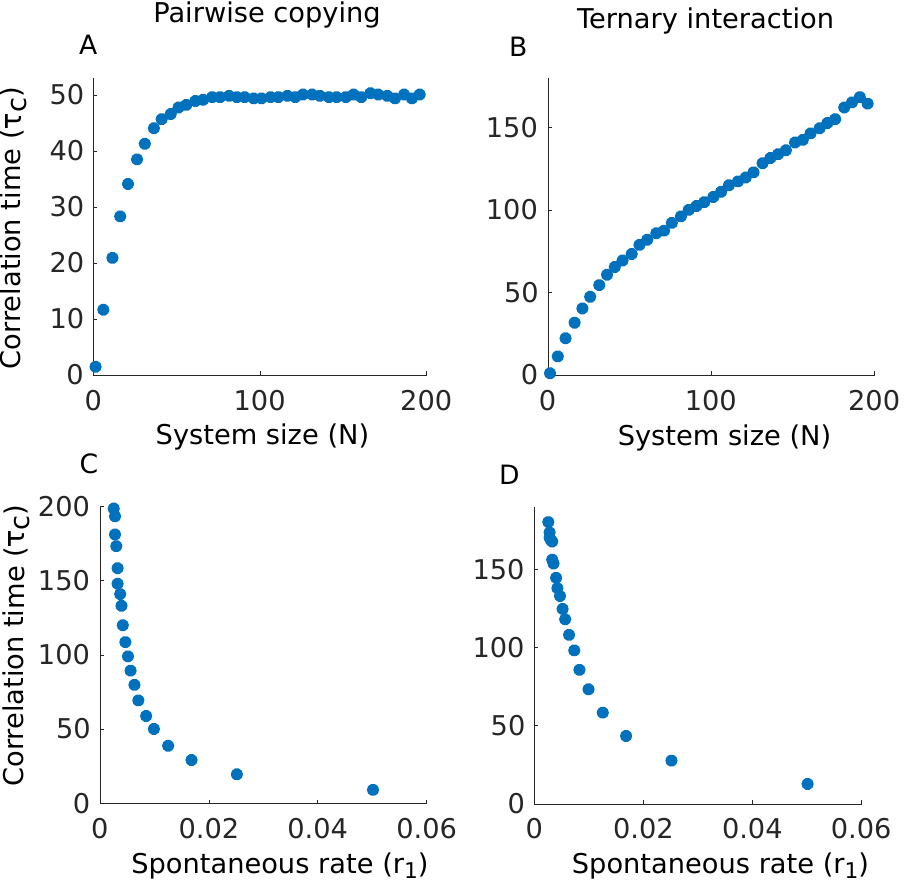}
\caption{\textbf{Correlation time of polarization as a function of system size ($N$) and spontaneous reaction rate ($r_1$) for both pairwise and ternary interaction model.} While the correlation time saturates with increasing $N$ for the pairwise interaction model {[\bf A]}, it increases monotonically with $N$ for the ternary interaction model {[\bf B]}. However with increasing $r_1$ the correlation time decreases exponentially for both the models {[\bf C and D]}}
\label{fig:tau_cont_var}
\end{figure*}

Using the above procedure we calculated correlation time $\tau_c$ for different system sizes for both the models. As shown in figure~\ref{fig:corr_fit}, the ACF decays exponentially with lag time. Using such  exponential fits we calculated the correlation times of time series for different values of system sizes ($N$) and the spontaneous reaction rate ($r_1$). We expected the correlation time to increase with increasing system size ($N$) and decreasing spontaneous reaction rate ($r_1$). This is because both of these, i.e. increasing $N$ or decreasing $r_1$, in general should reduce the strength of the intrinsic noise in the system. This reduction in the strength of noise should lead the system to reside close to a stable state for longer times, thus increasing the correlation time. As expected, we find that the correlation time increases with system size. For the pairwise copying model, however, it reaches an upper limit very quickly compared to the ternary interaction model (figure~\ref{fig:tau_cont_var} A \& B) where it gradually keeps increasing with system size (in the range of $N$ we considered). For smaller $r_1$, the correlation time is greater and reduces very sharply (almost exponentially) upon increasing $r_1$ for both the models (figure~\ref{fig:tau_cont_var} C \& D). Therefore we varied $r_1$ in our analysis to realize any changes in correlation time for the pairwise model. On the other hand, for completeness, we varied $N$ for the ternary model to realize changes in correlation time.

%figure Correlation time versus N figure 2
\section{Pairwise and ternary interaction models with $N=100$ and $N=200$}
We find the timescales to apply the method to reconstruct a stochastic differential equation from the data to be consistent for different values of system size ($N$) for both the pairwise and the ternary interaction models (figure~\ref{fig:param}).
\setcounter{figure}{0}
% SI Figure 3
\begin{figure*}[!]
\centering
\includegraphics[width=0.8\textwidth]{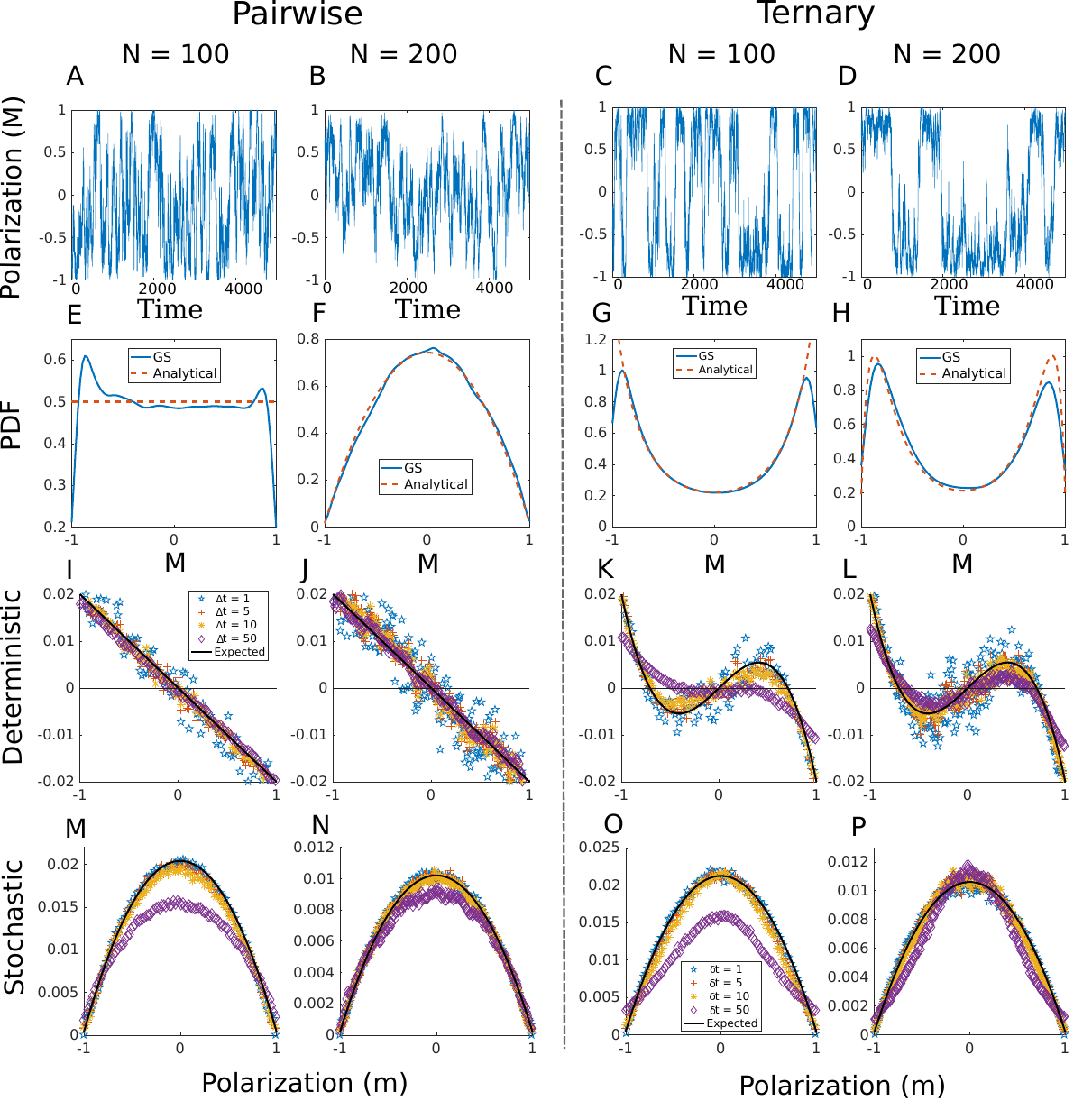}
\caption{\textbf{} Same as Figure 2 of main text with $N=100$ and $N=200$. These plots confirm that construction of deterministic forces is best for $\Delta t \le \tau_c$ and the optimum $\delta t$ is much smaller for stochastic forces. For both the models, the deterministic part ($F(m)$) of the SDE does not depend on $N$, however the stochastic part ($G^2(m)$) scales as $1/N$. Thus, the effect of stochasticity decreases with increasing system size - as expected from analytical calculations. For the pairwise copying model at $N = 100$, we find a flat probability distribution (corresponding to a critical value, see~\cite{biancalani2014PRL}); for $N = 200$, the distribution peaks at the disordered state $M = 0$. On the other hand, for the ternary interaction model, order due to deterministic forces does not depend on system size. Therefore, with increasing $N$ the modes in the distribution of $M$ become sharper at the deterministic stable fixed points.}
\label{fig:param}
\end{figure*}
\section{Consistency analysis}
Our analysis is based on the crucial assumption that the data of group dynamics can be modeled by a stochastic differential equation (SDE) of the form described by equation~1 of the main manuscript. According to this the noise in the dynamics is assumed to be Gaussian and uncorrelated. Since this is an assumption, its consistency needs to be checked against the data. We perform this analysis in two steps. First, we test the consistency of the method to extract the deterministic and stochastic functions by applying it on the the data generated from numerical integration the SDE (that incorporates such noise). Second, we compare the correlation function structure of this data with the one generated using the Gillespie simulations (GS) of the individual-based models.  
\subsection{Consistency of the reconstruction method}
By applying the reconstruction method to the data generated using SDE numerical integration we find that the time scales required to reconstruct the deterministic and the stochastic function are consistent with those for deriving the SDE from the data of the individual-based models. 

For the reconstruction of the deterministic part, a time scale considerably higher than the smallest time scale and of the order of correlation time (figure~\ref{fig:recon_sim_sde}A \&~\ref{fig:recon_sim_sde}B) needs to be used, i.e. $\Delta t \approx 50$. However for the stochastic part, the smallest time scale is the most appropriate for reconstruction (figure~\ref{fig:recon_sim_sde}C \&~\ref{fig:recon_sim_sde}D), i.e. $\delta t = 1$. Both these observations about the two time scales are consistent with SDE reconstruction from individual-based models data. 

%figure Reconstruction from simulations of SDE figure C1
\setcounter{figure}{0}
\begin{figure*}[!]
\centering
\includegraphics[width=0.6\textwidth]{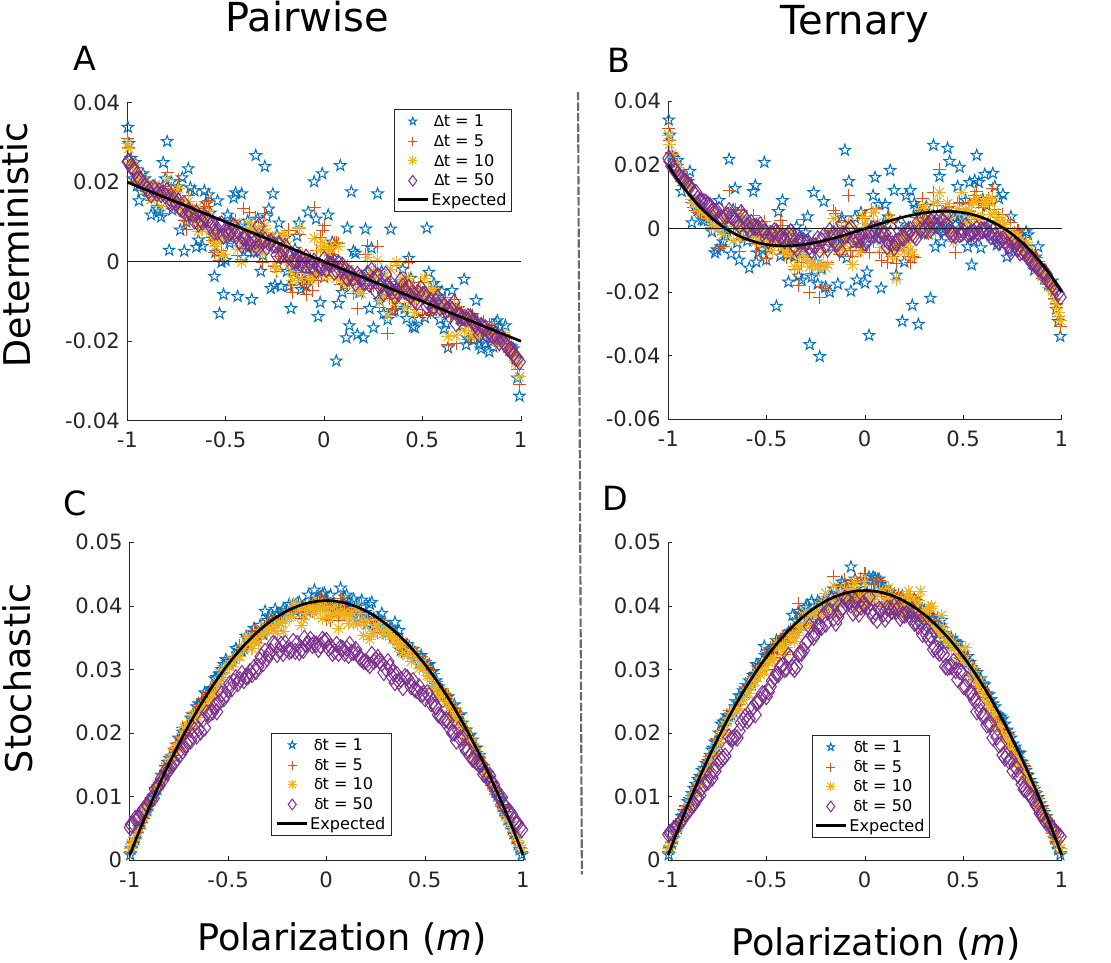}
\caption{\textbf{Reconstruction of functions from SDE simulations} Deterministic and stochastic function reconstructed from data generated from SDE simulations for validating the model against the data. [{\bf A and B}] Reconstructed deterministic function for different values of $\Delta t$ for the pairwise and the ternary models, respectively. [{\bf C and D}] Reconstructed stochastic function for different values of $\delta t$ for the pairwise and the ternary models, respectively. The time scales for reconstruction here match with those of the data from the individual based models (figure 2 of the main manuscript). Parameters used: same as figure 2 of the main manuscript.}
\label{fig:recon_sim_sde}
\end{figure*}
\subsection{Consistency of the SDE model}
To check the consistency of the SDE model certain features of the data generated from the SDE simulations and the Gillespie simulation (GS) of the individual-based models need to be compared. Here, we analyze the autocorrelation function (ACF) for this purpose. 

We find striking similarities between the ACF of both the data, i.e. from the GS and the SDE models (figure~\ref{fig:corr_fit}). This suggests that the SDE model well approximates the mesoscopic dynamics from the GS. \\

Overall, these consistency analyses suggest that the SDE model is a reasonable approximation to the dynamics of the individual-based models. Furthermore, the reconstruction method works equally well for both kinds of data. 

However, when working with real data of a group dynamics, additional tests may be necessary to ensure that SDE (equation~1 of the main manuscript) well describes the underlying dynamics. For example, one may need to test if the mesoscopic dynamics is indeed driven by Gaussian and uncorrelated  noise as shown in Section 3.3 of the main manuscript. 
\section{Underlying noise structure}
As analyzed in the main text, we find that the structure of the underlying noise remains Gaussian for coarser time scales as well (figure~\ref{fig:noise_coarse}). However, using a coarser time scale may lead to spurious calculations of noise as stochastic effects may average out for larger $\delta t$. 
%
%figure Noise structure D1
\setcounter{figure}{0}
\begin{figure*}[!]
\centering
\includegraphics[width=0.6\textwidth]{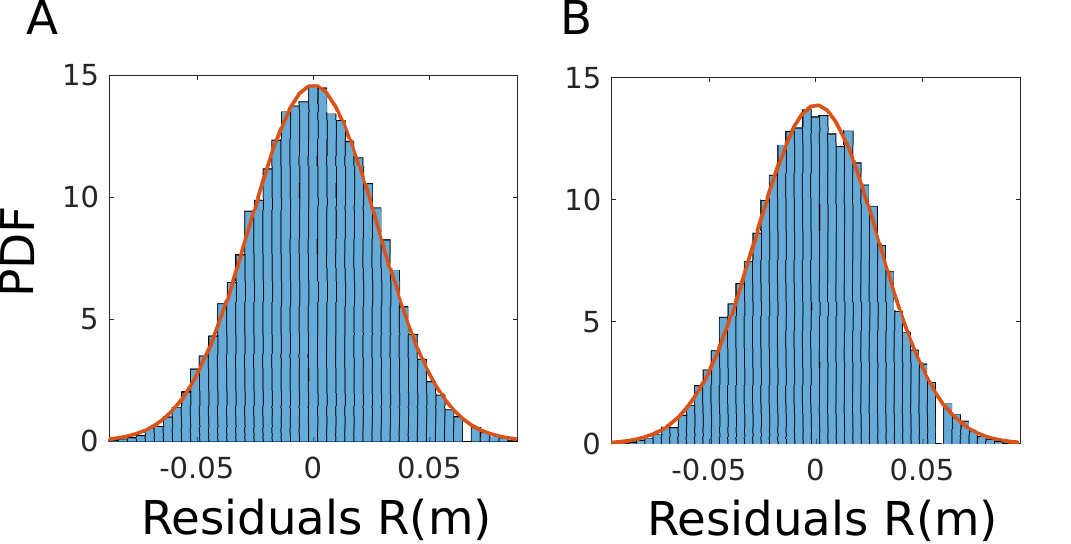}
\caption{\textbf{Noise structure for coarser time scales} For both the [{\bf A}] pairwise and the [{\bf B}] the ternary model we find that the noise structure remains Gaussian for coarser resolution ($\delta t = 100$) in the data.}
\label{fig:noise_coarse}
\end{figure*}

\end{document}